\documentclass[preprint2]{aastex}

\usepackage{graphicx}
\usepackage{subfigure}
\usepackage{amsmath}
\usepackage{amsfonts}
\usepackage{amssymb}
\usepackage[T1]{fontenc}
\usepackage{calligra}

\def\be{\begin{equation}}
\def\ee{\end{equation}}

\def\beq{\begin{eqnarray}}
\def\eeq{\end{eqnarray}}

\begin{document}

\title{All Fierz-Paulian massive gravity theories have  ghosts or superluminal modes} 

\author{Andrei Gruzinov}

\affil{CCPP, Physics Department, New York University, 4 Washington Place, New York, NY 10003}

\begin{abstract}

We show that all non-linear completions of the  Fierz-Pauli massive gravity are ruled out, because all theories which might be ghost-free have superluminal modes. 

~~

~~

\end{abstract}

\section {FP theories}

By Fieirz-Paulian massive gravity theories (FP) we mean all nonlinear completions of the original Fierz and Pauli (1939) linearized massive gravity. FP can be represented by 4 non-canonical scalars interacting with Einsteinian gravity ( Chamseddine and Mukhanov 2010,  Dubovsky 2004, Arkani-Hamed et al 2003). The action of the FP theory is
\beq
&S=S_{EH}+S_{FP}, &\\
&S_{EH}=-{1\over 2}\int d^4x\sqrt{-g}R,& \\
&S_{FP}=m^2\int d^4x\sqrt{-g}{\cal L}.&
\eeq
Here ${\cal L}$ is an arbitrary symmetric function of the eigenvalues of the matrix
\be
\label{H}
H^A_B=\eta _{BC}g^{\mu \nu}\partial _\mu\phi ^A\partial _\nu\phi ^C,
\ee
where $g_{\mu \nu}$ is the metric tensor, $\eta _{AB} \equiv {\rm diag} (1,-1,$ $-1,-1)$ is the Minkowski matrix, and $\phi ^A$, $A=0,1,2,3$,  are the 4 scalar fields.

De Rham et al (2010) show that all FP theories have the Boulware and Deser (1972) ghost, unless the theory belongs to a special 2-parameter family of FP theories -- call it FP2. In FP2 the 4-field Lagrangian is (Nieuwenhuizen 2011)
\be \label{dR}
{\cal L}=\sum \lambda _A\lambda_B+c_1\sum \lambda _A\lambda _B\lambda _C+c_2\lambda _0\lambda _1\lambda _2\lambda _3,
\ee
where the sums are over all all-distinct pairs and triples of indices, $\lambda _A$ are the four eigenvalues of the matrix
\be \label{k}
\delta ^A_B - \sqrt{ H^A_B },
\ee
and $c_1$, $c_2$ are the two dimensionless parameters of the theory.

It is not known whether the FP2 family is ghost-free, but we show here that it does not matter, because the scalar sector of  FP2 has superluminal modes in flat spacetime.  
\section {Superluminal modes in FP2}

Let $(t,x,y,z)$ be Minkowski coordinates in flat spacetime. Since our Lagrangian depends only on the derivatives of the fields, any 4 linear functions of coordinates solve the equations of motion, and can be chosen as background.  For our purposes, full generality is not needed. It is sufficient to consider the following background fields:
\be \label{bg}
\phi ^A = (t, x+\epsilon t, y, z).
\ee
We will ultimately take small $\epsilon$, because we are interested in the backgrounds close to  $\phi ^A = (t, x, y, z)$ -- a zero-stress-energy background of FP with $g_{\mu \nu}=\eta _{\mu \nu}$. But at this point of the calculation, $\epsilon$ is just an arbitrary constant. 

Consider infinitesimal (linear) perturbation of the form \footnote{It can be shown that linear perturbations of $\phi ^2$ propagating along $x$ on the background (\ref{bg}) do not excite perturbations of the fields $\phi ^0$, $\phi ^1$, and $\phi ^3$.}
\be \label{pt}
\delta \phi ^A = (0, 0, \chi (t,x), 0).
\ee
To second order in  $\chi$, that is for linear waves, the Lagrangian is (Appendix) \beq \label{lat}
{\cal L}= {1\over 2\sqrt{4-\epsilon ^2}}(\dot{\chi }^2-\chi '^2)~~\nonumber \\
+{c_1+1-1/\sqrt{4-\epsilon ^2}\over 2+\sqrt{4-\epsilon ^2}}(\epsilon\dot{\chi }\chi '-{1\over 2}\epsilon^2\chi'^2).
\eeq

To leading order in $\epsilon$, 
\be
{\cal L}= {1\over 4}(\dot{\chi }^2-\chi '^2)+{2c_1+1\over 8}\epsilon\dot{\chi }\chi ',
\ee
giving the phase speed $v=\pm 1+{2c_1+1\over 4}\epsilon$, which is superluminal unless $c_1=-{1\over 2}$. It remains to consider the case $c_1=-{1\over 2}$. For $c_1=-{1\over 2}$, to third order in $\epsilon$,
 ${\cal L}\propto \dot{\chi }^2-\chi '^2-{1\over 16}\epsilon ^3\dot{\chi }\chi '$, which is superluminal.

\section*{Acknowledgments}

I thank S. Dubovsky, G. Gabadadze, M. Mirbabayi, V. Mukhanov for useful discussions.

\renewcommand{\theequation}{A\arabic{equation}}
\setcounter{equation}{0} 

\section*{Appendix}

With $g_{\mu \nu }=\eta _{\mu \nu }$, and the scalar fields given by (\ref{bg}, \ref{pt}), the matrix $H^A_B=\eta _{BC}g^{\mu \nu}\partial _\mu\phi ^A\partial _\nu\phi ^C$ is 
\be 
 H^A_B = \left ( \begin{array}{llll}

1             & -\epsilon                           & -\dot{\chi }                         & 0 \\
\epsilon  & 1-\epsilon ^2                    & -\epsilon \dot{\chi }+\chi '  & 0 \\
\dot{\chi }&  -\epsilon \dot{\chi }+\chi ' &1-\dot{\chi }^2+\chi '^2       & 0 \\
0             &  0                                      & 0                                       & 1 
 
\end{array}\right ) .
\ee
We need to calculate the eigenvalues $\lambda _A$ of the matrix $\delta ^A_B - \sqrt{ H^A_B }$. We first calculate $\kappa _A$ -- the eigenvalues of $H$. From 
\be
{\rm det} (H-\kappa )=\Pi (\kappa _A-\kappa)
\ee
we get
\be \label{k3}
\kappa _3=1,
\ee
\be \label{kp}
\kappa _0\kappa _1\kappa _2=1,
\ee
\be \label{ks}
\kappa _0+\kappa _1+\kappa _2=t_1,
\ee
\be \label{kp2}
\kappa _0\kappa _1+\kappa _0\kappa _2+\kappa _1\kappa _2=t_2,
\ee
where we denoted
\be
t_1=3-\epsilon^2-\dot{\chi }^2+\chi '^2,
\ee
\be
t_2=t_1-\epsilon ^2\chi '^2+2\epsilon \dot{\chi }\chi '.
\ee

From (\ref{k3}), $\lambda _3=0$, and the Lagrangian simplifies to 
\be
{\cal L}= (\lambda _0\lambda _1+\lambda _0\lambda _2+\lambda _1\lambda _2)+c_1\lambda _0\lambda _1\lambda _2
\ee
Here $\lambda _A =1-\kappa _A^{1/2}$, giving
\be \label{la}
{\cal L}= 3-2s_1+s_2+c_1(s_2-s_1),
\ee
where
\be
s_1=\kappa _0^{1/2}+\kappa _1^{1/2}+\kappa _2^{1/2},
\ee
\be
s_2=\kappa _0^{1/2}\kappa _1^{1/2}+\kappa _0^{1/2}\kappa _2^{1/2}+\kappa _1^{1/2}\kappa _2^{1/2}.
\ee
From (\ref{kp}, \ref{ks}, \ref{kp2}) we have
\be \label{s}
s_1^2-2s_2=t_1, ~~ s_2^2-2s_1=t_2,
\ee

To zero order in $\chi$, (\ref{s}) gives 
\be
s_1=s_2=s\equiv 1+\sqrt{4-\epsilon ^2}.
\ee
Then we define
\be
s_1=s+\delta s_1, ~~s_2=s+\delta s_2. 
\ee
To second order in $\chi$, (\ref{s}) gives
\be
2s\delta s_1-2\delta s_2=-\dot{\chi }^2+\chi '^2,
\ee
\be
2s\delta s_2-2\delta s_2=-\dot{\chi }^2+\chi '^2-\epsilon ^2\chi '^2+2\epsilon \dot{\chi }\chi '.
\ee
This system gives $s_1$ and $s_2$. We use the resulting expressions in (\ref{la}) and get (\ref{lat}).

\end{document}